# 6 MV photon beam modeling for Varian Clinac iX using GEANT4 virtual jaw

Byung Yong Kim, Hyung Dong Kim, Dong Ho Kim, Jong Geun Baek and Su Ho Moon

*Department of Physics, Yeungnam University, Gyeongsan 712-749, Korea*

Gwang Won Rho

*Department of Radiation Oncology, Sunlin Medical Center, Pohang 791-704, Korea*

Jeong Ku Kang

*Department of Radiation Oncology, Presbyterian Medical Center, Jeonju 560-750, Korea*

Sung Kyu Kim[*]

*Department of Radiation Oncology, Yeungnam University College of Medicine, Daegu 705-717, Korea*

Most virtual source models (VSM) use beam modeling, with the exception of the patient-dependent secondary collimator (jaw). Unlike other components of the treatment head, the jaw absorbs many photons generated by the bremsstrahlung, which decreases the efficiency of the simulation. In the present study, a new method of beam modeling using a virtual jaw was applied to improve the calculation efficiency of VSM. The results for the percentage depth dose and profile of the virtual jaw VSM calculated in a homogeneous water phantom agreed with the measurement results for the CC13 cylinder-type ion chamber within an error rate of 2%, and the 80–20% penumbra width agreed with the measurement results within an error of 0.6 mm. Compared with the existing VSM, in which a great number of photons are absorbed, the calculation efficiency of the VSM using the virtual jaw was expected to increase by approximately 67%.

Email: skkim@med.yu.ac.kr

Fax: +82-53-624-3599

# I. INTRODUCTION

Various components of the treatment head complexly interact with electrons and photons. Therefore, in most Monte Carlo (MC) simulations performed in the dose calculation for patients, the beam modeling for the treatment head takes precedence to increase the efficiency of the simulation process and the accuracy of the result [1-4]. The virtual source model (VSM) is a method of beam modeling that reconstructs the treatment head using several virtual beam sources [5, 6, 7]. In VSM, the virtual beam source is designed using the phase space (PS) data, which are the simulation results of the treatment head, along with the percentage depth dose (PDD) and profile, which are measured results [7, 8, 9]. Because the secondary collimator (jaw) is a component of the treatment head that is a patient-dependent variable, most MC simulations that use a VSM simulate the interaction process with particles in the jaw [10, 11, 12]. Most of the particles that pass through the jaw are absorbed by the jaw and do not contribute to the patient's dose. If the particles that are eventually absorbed in the jaw are not generated, then the efficiency of the MC simulation will be greatly improved while the same dose is delivered to the patient.

The jaw determines the size of the radiation field and forms the penumbra region. Furthermore, it produces scattered photons to affect the beam fluence. We used a virtual jaw to recreate two types of effects caused by the jaw. The photon beam generated from the treatment head can be divided into a primary photon beam and extrafocal photon beam, depending on its characteristics [13, 14]. We defined the primary photon beam source as a circle and limited the primary photon beam's direction of propagation to the radiation field in order to reveal the shielding effects of the jaw. Moreover, we included the effect of the jaw as a beam source, which is known to be <1%, in the extrafocal photon beam [6, 15]. As a result, the entire treatment head, including the jaw, was implemented with a VSM, and the occurrence of photon beams that do not contribute to the patient region was minimized. The goals of this study are to develop a virtual jaw VSM that can minimize the absorption of particles in the treatment head and to assess the accuracy and efficiency of the MC simulation.

## II. MATERIALS AND METHODS

1. Beam modeling

Figure 1 shows the beam modeling process of the virtual jaw VSM. We first measured the PDD and profile of Varian Clinac iX 6 MV photons in distance intervals of 1 mm in a homogeneous water phantom using a CC13 (IBA dosimetry, German) chamber. In addition, to collect the PS data, we implemented a direct MC simulation and compared its results with the measured values. On the basis of the PS data and the measured values of the PDD and profile, we designed the primary photon beam source and extrafocal photon beam source of the virtual jaw VSM and determined the values of various factors. The virtual jaw VSM produced in this manner calculated the dose distribution for field sizes of $10 \times 10$ cm$^2$ and $20 \times 20$ cm$^2$ in a homogeneous water phantom and compared the PDD and profile with the measured values. We determined the final values of the factors through repeated comparisons of the calculation results with the measurement results.

For the direct MC simulation, we used the Geant4 code to reproduce the treatment head of Varian Clinac ix 6 MV as closely as possible. Figure 2(a) shows a schematic of the direct MC simulation, which sends the electrons to the target and then simulates the process of complex interactions created by the various components. For the field sizes of $10 \times 10$, $20 \times 20$, $30 \times 30$, and $40 \times 40$ cm$^2$, $2.4 \times 10^{10}$ incident electrons were generated, from which the dose distribution was calculated. The size of the water phantom used in the direct MC simulation was $50 \times 50 \times 50$ cm$^3$, and the voxel size was $0.5 \times 0.5 \times 0.5$ cm$^3$. To test the validity of the direct MC simulation, the calculated PDD and profile were compared with the measured values. The PS data were collected from PS27 and PS45, which are located 27 and 45 cm, respectively, from the target.

The virtual jaw VSM consisted of a primary photon beam source and an extrafocal photon beam source, divided on the basis of the measured values and PS data (Fig. 2(b)). Physical penumbra can be divided into geometric penumbra and transmission penumbra [16]. We used the generation principle of geometric penumbra to determine the appropriate primary photon beam source radius for displaying physical penumbra. The primary photon beam source is a circular virtual source with a radius of 3.1 mm. The direction of the primary photon beam was pointed from an arbitrary point of the source to an arbitrary point of the aperture defined by the jaw [17]. Therefore, the primary photon beam directed toward the jaw is not generated. However, because the shape of the primary photon beam source is circular, with a radius of 3.1 mm, a penumbra region approximately 7.6 mm in size, is formed on the surface at an SSD of 100 cm, as shown in Fig. 2(b). The extrafocal photon beam source was situated 45 cm from the target, with the same shape as the radiation field, to achieve a Gaussian distribution [18, 19]. The size of the water phantom used in the dose calculation of the virtual jaw VSM was $50 \times 50 \times 50$ cm$^3$, and the voxel size was $0.1 \times 0.1 \times 0.5$ cm$^3$. In the field sizes of $10 \times 10$ and $20 \times 20$ cm$^2$, beams of $1 \times 10^9$ and $4 \times 10^9$ were generated, respectively, to calculate the dose distribution. The factors, such as the energy spectrum, standard deviation, source weight, and fluence weight, were calculated from the results of the PS analysis, and the final values were determined through repeated comparisons of the calculation results with the measurement results.

2. Beam modeling evaluation

The dose calculation results of the direct MC simulation and virtual jaw VSM were compared with the measurement results obtained using the CC13 chamber, and the differences in the dose distribution were evaluated. The direct MC simulation calculated the PDD and profile in the fields with sizes of $10 \times 10$, $20 \times 20$, $30 \times 30$, and $40 \times 40$ cm$^2$, and the VSM calculated the PDD and profile in the fields with sizes of $10 \times 10$ and $20 \times 20$ cm$^2$. The two beam models calculated the profiles at the depths of 1.5, 5, 10, and 20 cm. The PDD and profile calculation results from the virtual jaw VSM are shown in 0.1 cm intervals, as are the measurement results. For the dose at each point, the averages of the dose values scored by the neighboring 25 (5 × 5 × 1) voxels were used, and the total volume of the voxels used was 0.125 cm$^3$ (0.5 × 0.5 × 0.5 cm$^3$). To evaluate the penumbra region of the VSM, we calculated the 20–80% penumbra width at each profile depth for the +x and –x regions and compared them with the measurement results.

In the MC simulation, the efficiency ($\varepsilon$) was expressed as $\varepsilon = 1/s^2 T$ and T was nearly proportional to N [5, 20]. Here, $s^2$ is the variance ($\sigma^2$), T is the CPU time, and N is the number of histories simulated. In the direct MC simulation, the two PSs can be regarded as beam sources for a new beam model that uses each of the PS data. Therefore, the differences in the number of particles collected in the two PSs are proportional to the differences in the calculation time [20], and the jaw can be viewed as the cause of these differences. Although the circular radiation field generated by the target in the treatment head is consistent regardless of the field size, the radius of the circular radiation field can be adjusted in the beam model to suit the application. Therefore, in a beam model that uses a jaw for shielding, the minimum shielded area would occur when the initial circular radiation field is changed to a square radiation area that inscribes the circle. Accordingly, the minimal efficiency increase was evaluated by comparing the number of particles and the radiation areas between the case of using a jaw to minimize shielding and the case with the same results but without any shielding with the use of a jaw. The beam model that has PS27 as the beam source can be considered for the case with shielding from a jaw, and the beam model that has PS45 as the beam source can be considered for the case that does not require a jaw. These two beam models can be assumed to have the conditions that influence the calculation time, except for the number of particles. The relative increases in the efficiency for the field sizes of 10 × 10, 20 × 20, and 30 × 30 cm$^2$ were calculated, and the field size of 40 × 40 cm$^2$ was excluded in the calculation because it contains an area that is outside the radiation angle of the primary collimator.

### III. RESULTS AND DISCUSSION

The calculated PDD and profile results from the direct MC simulation exhibited <3% error and uncertainty with respect to the measurement results from the regions within the field. Figures 3 and 4 indicate the comparisons between the calculated PDD and profile results from the virtual jaw VSM and the measurement results. When the particles of $1 \times 10^9$ and $4 \times 10^9$ were irradiated in the field sizes of $10 \times 10$ and $20 \times 20$ cm$^2$, respectively, the calculated PDD and profile results of the virtual jaw exhibited <2% error with respect to the measurement results from the regions within the field, and the uncertainty was <1.5%. Further decreases in the error and uncertainty rates are expected if more accurate factors are applied and the number of incident particles is increased.

In the comparison of the 80–20% penumbra width in Fig. 5, the calculation results from the virtual jaw VSM exhibited a difference of <0.6 mm compared with the measurement results. At a depth of 1.5 cm, the 80–20% penumbra width calculation results from the virtual jaw VSM were smaller by 0.5 mm, on average, than the measurement results, and as the depth increased, the differences with respect to the measurement results decreased. At a depth of 20 cm, the results from the virtual jaw VSM were greater by 0.2 mm, on average, than the measurement results. The values from both the measurement and calculation for 80–20% of the penumbra width were approximately 6–12 mm, and a 60% change in the dose occurred. Therefore, considering that a 1% change in the dose leads to an error of approximately 0.1–0.2 mm, the effect of the virtual jaw can be viewed as highly accurate.

Table 1 indicates the irradiated area and the number of particles with respect to the field size for the case with shielding of a minimal area using the jaw and the case that does not require shielding by the jaw. Here, the percentage of the shielded area is a value obtained by dividing the area shielded by the jaw by the area of the circular radiation field. The percentage of the area shielded by the jaw is constant at 36.3%, regardless of the field size. The number of PS45 particles was a value that represented an approximately 40% reduction from the number of PS27 particles and exhibited a slight differences according to the field size. Therefore, when the number of particles absorbed by the jaw is reduced, it is possible to reduce the calculation time by approximately 40% or increase the efficiency by approximately 67%. For the virtual jaw VSM, the virtual jaw was applied by controlling the factor values without implementing any additional calculation processes. Thus, it is anticipated that the efficiency of the virtual jaw VSM will increase by approximately 67% compared with the conventional VSM that uses shielding from the jaw. The reduction in the rate of the number of particles was approximately 10% higher than the rate of the shielded area, which is attributed to the increase in the beam fluence due to the increase in the radius from the central axis, arising from the effects of a flattening filter [6, 21]. Therefore, because the jaw has the characteristic of shielding from the edges of the radiation field, the rate of reduction in the number of particles appears larger than the rate of the shielded area.

## IV. CONCLUSIONS

We developed a new VSM using a virtual jaw to reduce the additional calculation process that occurs in the jaw structure during the MC simulation. A comparison of the PDD and profile results between a virtual jaw VSM simulation and actual measurements indicated good agreement, with an error rate within 2%. Particularly, in the comparison of 80–20% penumbra width, which is greatly influenced by the jaw and target, because the error rate was within 0.6 mm with respect to the measurement results, it was determined that effects of the virtual jaw were well reflected. When the particles that would eventually be absorbed in the jaw were not generated, approximately 40% reduction in the calculation time and approximately 67% increase in the efficiency were observed. The efficiency of the virtual jaw VSM in delivering most of the generated particles to the phantom without loss is expected to be the same.

Table 1. Comparison of the radiation field and number of particles when the circular radiation field is shielded by the jaw to form a square radiation field that inscribes the circle

| Field size (cm$^2$) | Radiation Field | | | Number of Particles | | |
|---|---|---|---|---|---|---|
| | Area of circular radiation field (cm$^2$) | Area of square radiation field (cm$^2$) | Shielded area rate (%) | PS27 | PS45 | Rate of reduction in the number of particles (%) |
| 10 × 10 | 157.08 | 100 | 36.3 | $6.66 \times 10^7$ | $3.95 \times 10^7$ | 40.8 |
| 20 × 20 | 628.32 | 400 | 36.3 | $2.82 \times 10^8$ | $1.71 \times 10^8$ | 39.3 |
| 30 × 30 | 1413.72 | 900 | 36.3 | $6.69 \times 10^8$ | $4.04 \times 10^8$ | 39.7 |

Figure Captions.

Fig. 1. Beam modeling process of the VSM that uses the virtual jaw

Fig. 2. Schematic of (a) direct MC simulation and (b) virtual jaw VSM, reproduced by Geant4 code

Fig. 3. Comparison of PDD calculation result from virtual jaw VSM and PDD measurement results from Varian Clinac iX 6 MV for the field sizes of $10 \times 10$ and $20 \times 20$ cm$^2$ in the water phantom

Fig. 4. Comparison of profile calculation result from virtual jaw VSM and profile measurement results from Varian Clinac ix 6 MV for the field sizes of $10 \times 10$ and $20 \times 20$ cm$^2$ in the water phantom; (a) 1.5 cm depth, (b) 5 cm depth, (c) 10 cm depth, (d) 20 cm depth

Fig. 5. Comparison of 20−80% penumbra width calculated from the profile of the virtual jaw VSM and the measured profile. The minus and plus signs represent −x and +x areas of 20–80% penumbra width, respectively.

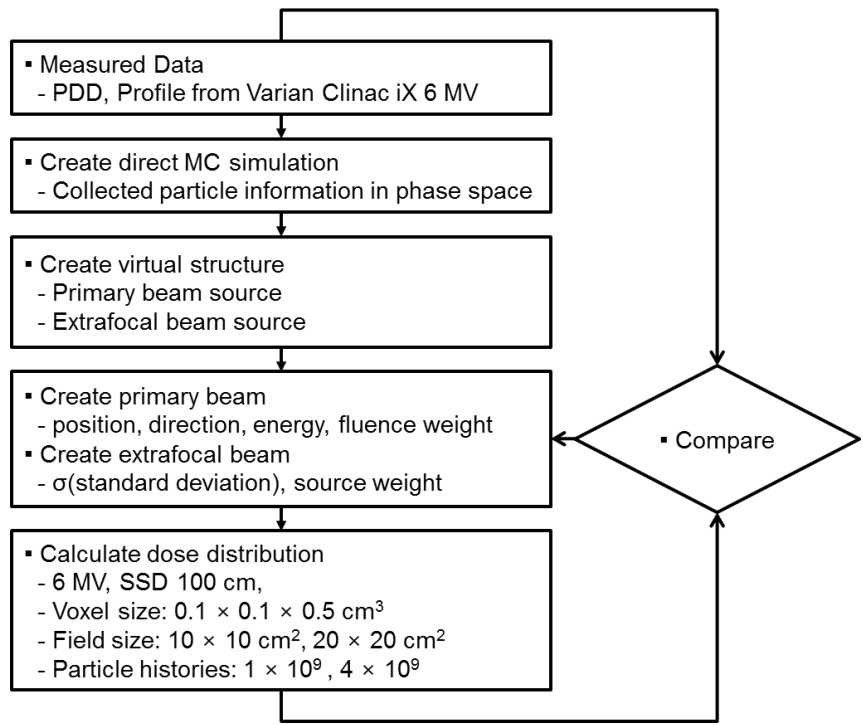

Fig. 1.

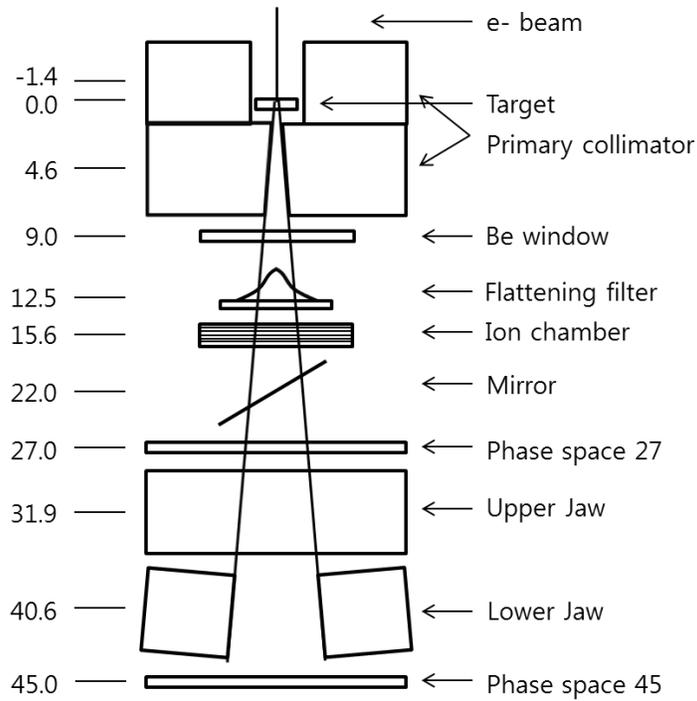

Fig. 2. (a)

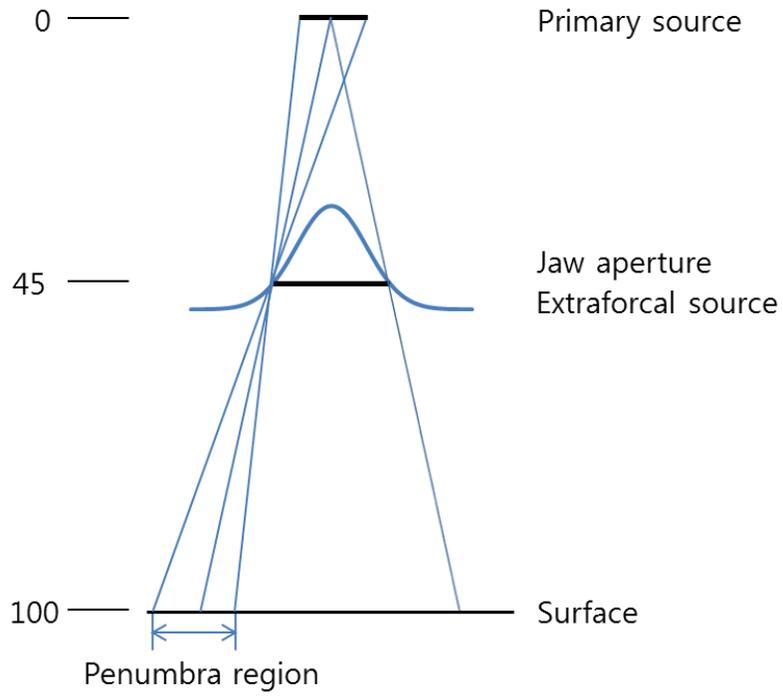

Fig. 2. (b)

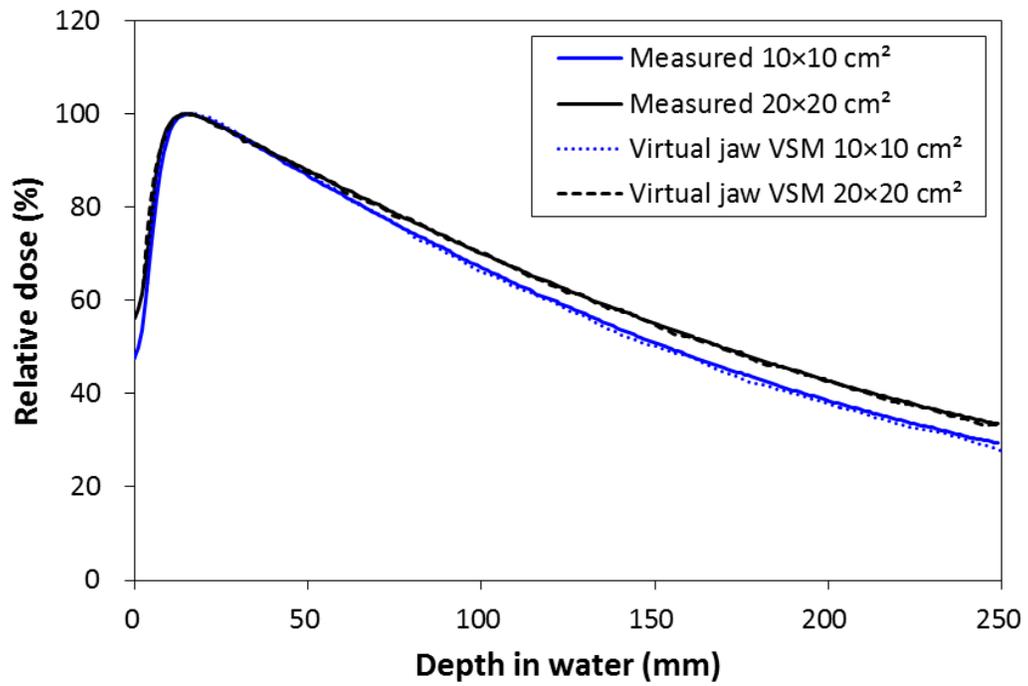

Fig. 3.

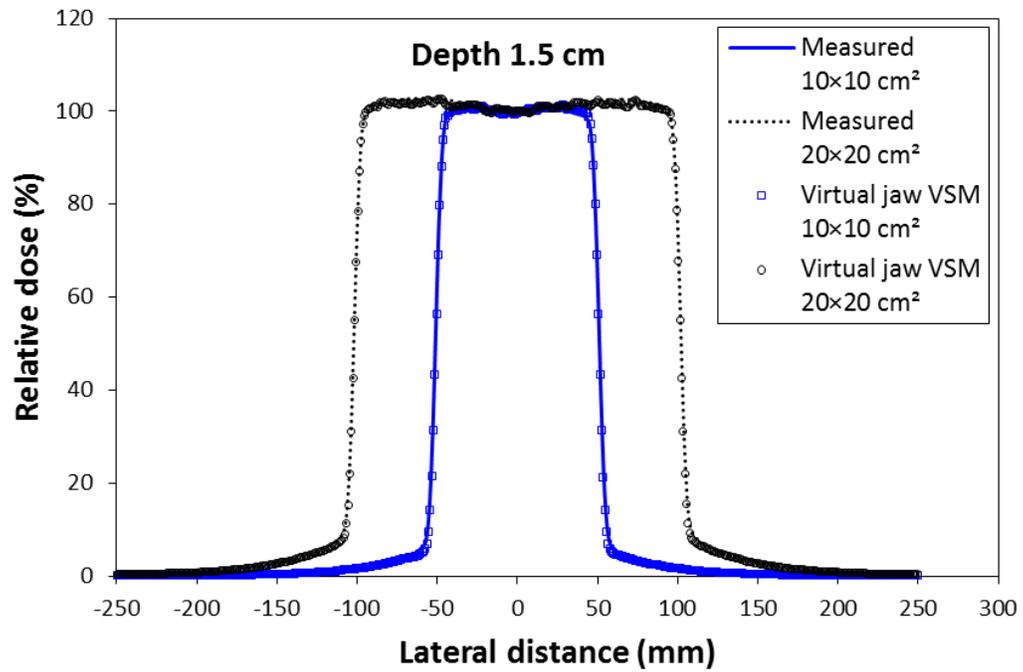

Fig. 4. (a)

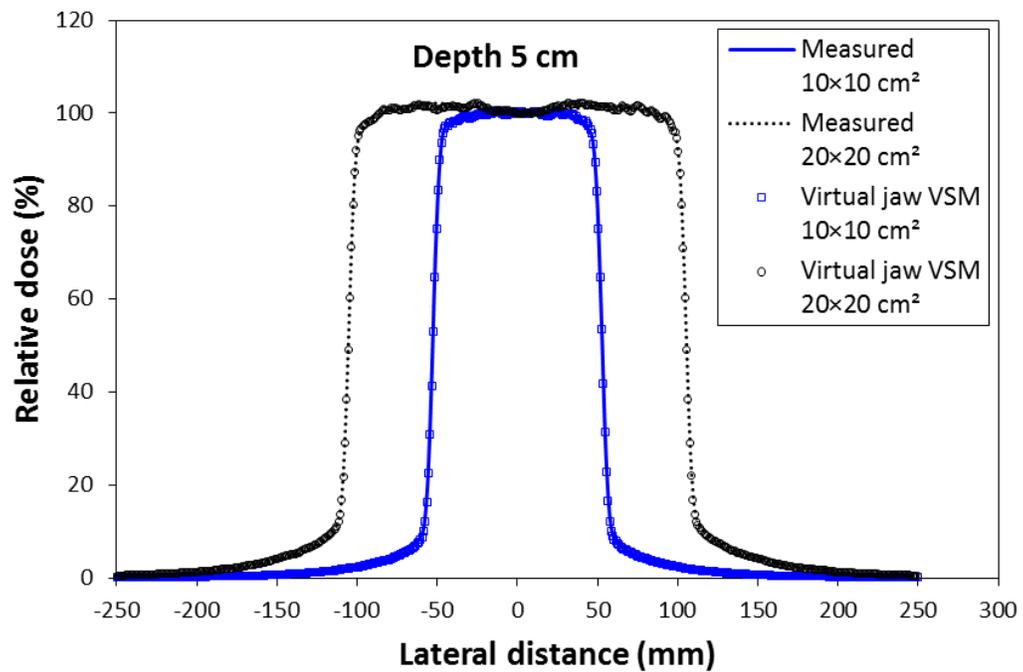

Fig. 4. (b)

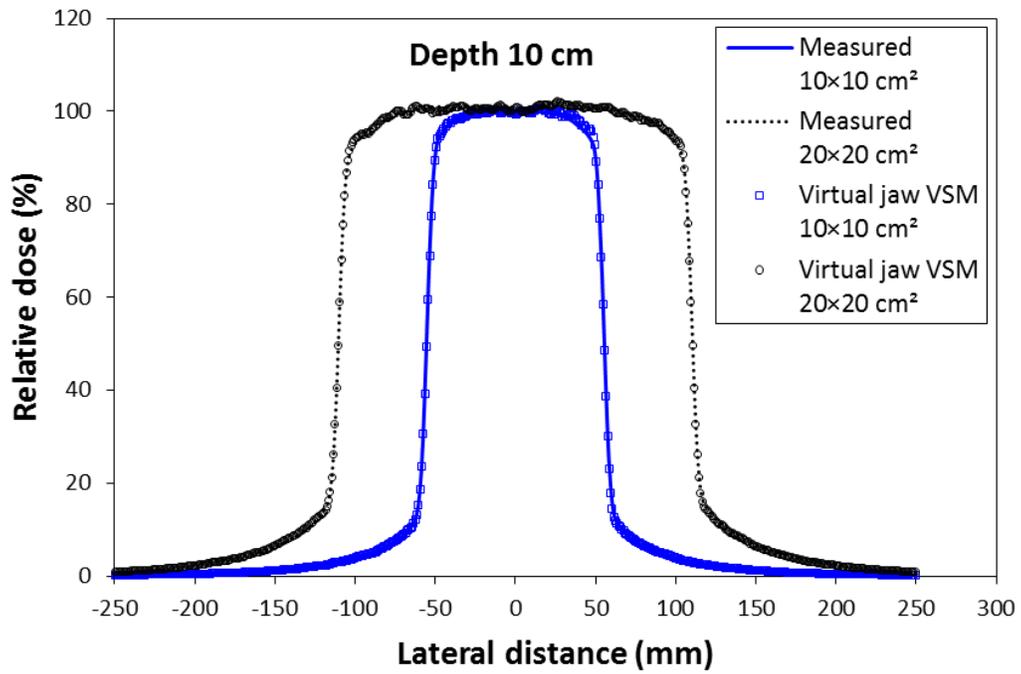

Fig. 4. (c)

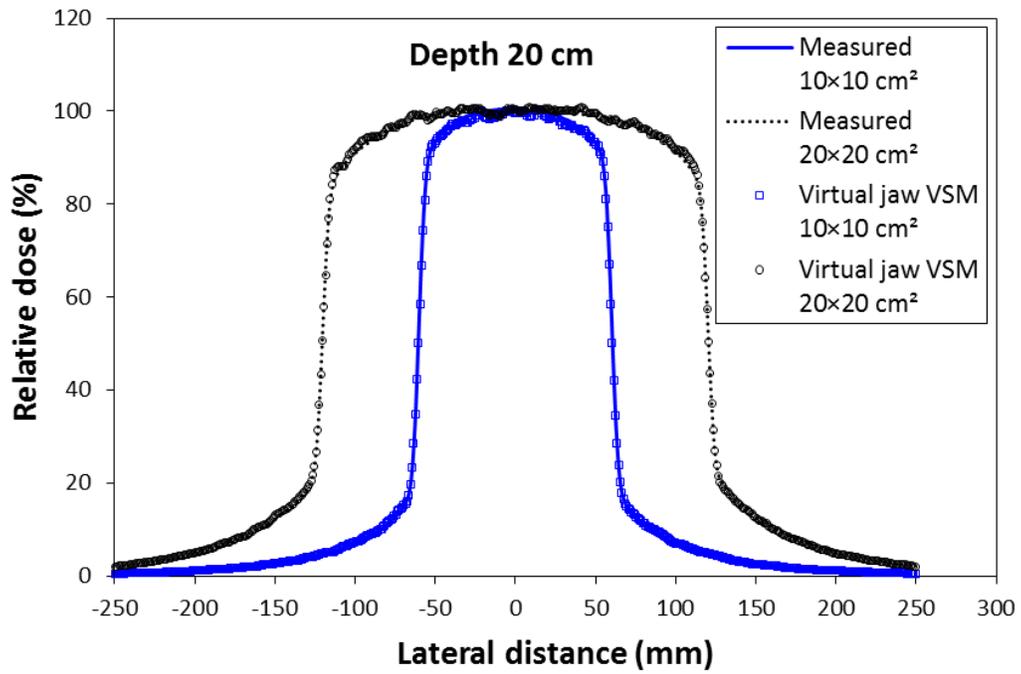

Fig. 4.

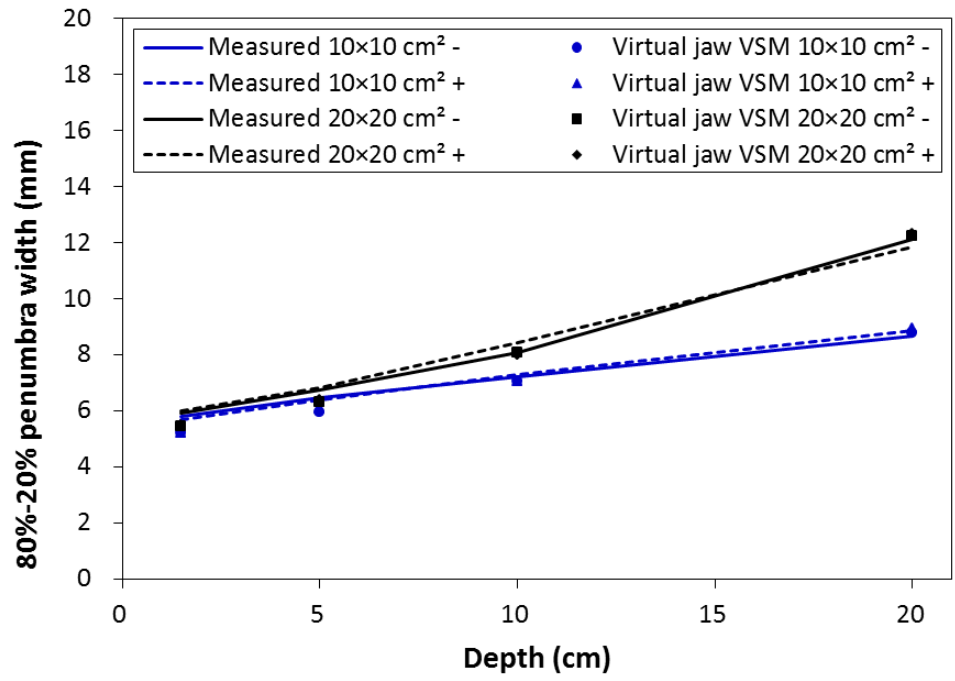

Fig. 5.